\RequirePackage{fix-cm}
\documentclass{svjour3}     
\smartqed  
\usepackage{graphicx}
\usepackage{latexsym}
\usepackage{hyperref}
\usepackage{ dsfont }
\usepackage{amsmath }
\usepackage{lineno,hyperref}
\usepackage{ bbold }
\usepackage{ amssymb }
\usepackage{epstopdf}
\usepackage{breqn}
\usepackage{import}
\usepackage{setspace}
\usepackage{ amssymb }
\usepackage{float}
\usepackage{subcaption}
\usepackage[numbers,sort&compress]{natbib}
 \modulolinenumbers[5]
\usepackage[inner=2.5cm,outer=2.5cm]{geometry}

\journalname{Nonlinear Dynamics}
\begin{document}

\title{Nonlinear vibration localisation in a symmetric system of two coupled beams}
%



\author{\mbox{Filipe Fontanela} \and
	\mbox{Alessandra Vizzaccaro*} \and
	\mbox{Jeanne Auvray} \and
	\mbox{Bj{\"o}rn Niederges{\"a}{\ss}}\and
	\mbox{Aur\'elien Grolet} \and
	\mbox{Lo{\"i}c Salles} \and
	\mbox{Norbert Hoffmann}
}


\institute{*Corresponding author: A. Vizzaccaro \\
\email{a.vizzaccaro17@imperial.ac.uk}
\\ \\
F. Fontanela \and A. Vizzaccaro \and L. Salles \and N. Hoffmann
	\at
	Imperial College London, Exhibition Road, SW7 2AZ London, UK 
	\and
	J. Auvray  
	\at
	Ecole Centrale Marseille, 13451 Marseille Cedex 20, France
	\and
	A. Grolet
	\at
	Ecole Nationale Superieure d'Arts et M\'etiers ParisTech, 59000 Lille, France
	\and 
	B. Niederges{\"a}{\ss} \and N. Hoffman
	\at
	Hamburg University of Technology, 21073 Hamburg, Germany
}

\maketitle
\begin{abstract}
We report nonlinear vibration localisation in a system of two symmetric weakly coupled nonlinear oscillators. A two degree-of-freedom model with piecewise linear stiffness shows bifurcations to localised solutions. An experimental investigation employing two weakly coupled beams touching against stoppers for large vibration amplitudes confirms the nonlinear localisation.
\end{abstract}
\section{Introduction}\label{intro}
The emergence of localised vibration in symmetric structures is a challenging problem in the aerospace industry due to high cycle fatigue \cite{whitehead1966effect, cornwell1989localization, judge2000experimental}. Usually, aerospace structures such as bladed-disks, antennas, and reflectors are composed of ideally identical substructures assembled in a symmetric configuration. In the linear regime, localisation may arise due to structural inhomogeneities originating in the manufacturing process or due to wear \cite{bendiksen1987mode, bendiksen2000localization}. In the aerospace industry, especially in the bladed-disk community, the phenomenon is thus widely referred to as a mistuning problem. The topic has attracted considerable attention in the literature, and research has mainly focused on effective numerical tools for prediction, experimental investigation, and the use of intentional mistuning during design stages \cite{hodges1982confinement, hodges1983vibration, castanier2006modeling, vargiu2011reduced, capiez2015mistuning}. \\
However, in some cases, due to inherent nonlinear phenomena, the assumption of linear vibration might be misleading. In the case of structural dynamics, nonlinearity may arise e.g. due to friction induced by internal joints, or vibro-impacts \cite{krack2017vibration}. It is also known that the emergence of localised vibration might be provoked by nonlinear effects, as an alternative to the linear localisation mechanisms in mistuning \cite{vakakis1992dynamics}. For example, even perfectly symmetric structures may experience localised vibrations due to the dependence of mode shapes on amplitude, or due to bifurcations. However, most of the available knowledge on this kind of nonlinear vibration localisation relies on results from minimal models, and only few experimental studies have attempted to demonstrate the existence of localised vibrations in symmetric structures due to nonlinear interactions \cite{emacinayfeh1997,sato2003observation}. \\
This paper thus reports the existence of nonlinear localised vibrations in a symmetric mechanical structure due to the presence of clearance nonlinearity. First we introduce a conceptual model with two degrees of freedom under the effect of a harmonically moving base. The bilateral contact phenomenon is assumed perfectly elastic, leading to a nonlinear mathematical model which is piecewise linear \cite{Reboucas2018}. In the free case, a nonlinear modal analysis is carried out, and we demonstrate that localised states bifurcate from the homogeneous out-of-phase mode. The results are similar to bifurcated states calculated in smooth systems, such as in chains of Duffing oscillators \cite{hillbackbones2015,Ikeda2013,papangelo2019multistability}. In the case of externally driven vibration, if the excitation is perfectly in phase, three kinds of stable response states may result. First, a purely linear configuration, where both masses vibrate in low amplitude and in phase. Second, a nonlinear configuration where both masses vibrate in large amplitude and in phase. And third, just one oscillator, either the first or the second, vibrates in large amplitudes, while the other one vibrates in small amplitudes and out of phase. An experimental validation of the numerical findings, based on a test-rig composed of two weakly coupled cantilever beams touching stoppers for large amplitude vibration, is reported. \\
The paper is organised as follows. In Sec. \ref{Sec:Theory} the numerical model is described and the outcomes of a non-linear modal analysis and a response analysis are presented. Section \ref{Sec:Exps} introduces the experimental test used to validate the results from the model. Sec. \ref{Sec:Conc} discusses the main findings and suggests directions for future investigation.
\section{Numerical model} \label{Sec:Theory}
\subsection{Description of the model}
\begin{figure}
		\centering
		\vspace{-0.1cm}
		\includegraphics
		[scale=.5]{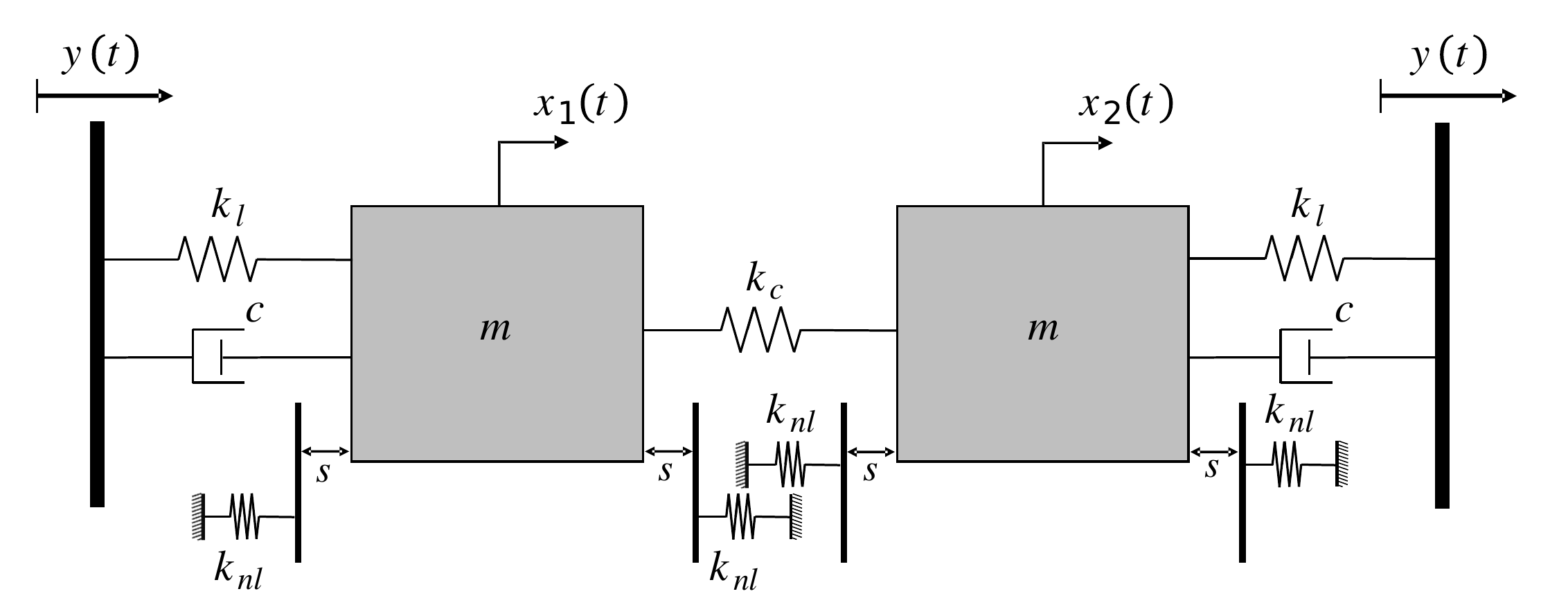}
		\caption{Symmetric system under investigation.}
		\label{Fig:PhySys}
\end{figure}

\begin{figure}
\centering
\begin{subfigure}{.5\textwidth}
  \caption{\hspace{3.5cm} }
\centering
\vspace{-0.25cm}
  \includegraphics[width=.8\linewidth]{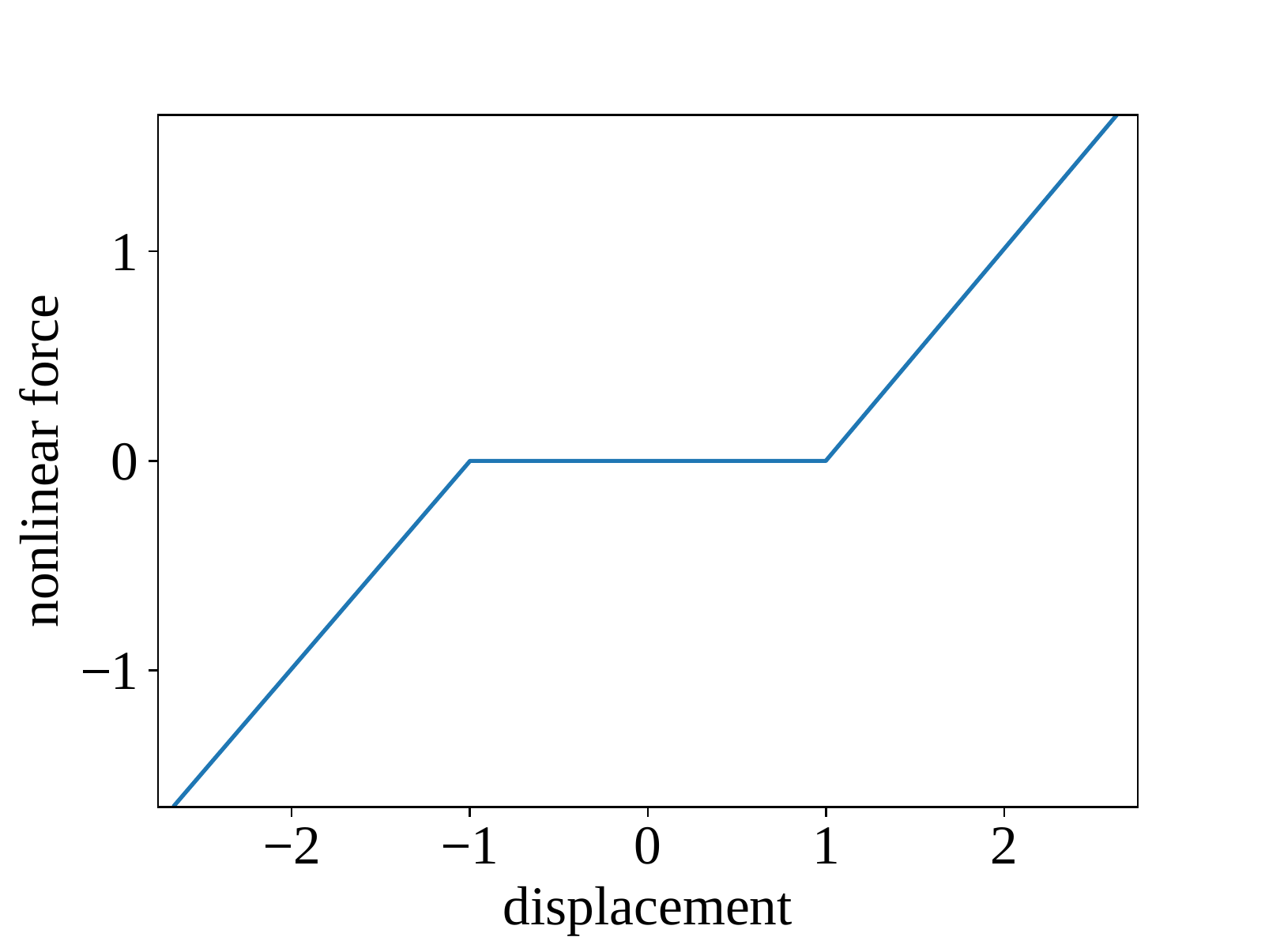}
  \label{fig:sub1}
\end{subfigure}%
\begin{subfigure}{.5\textwidth}
 \caption{\hspace{3.5cm} }
  \vspace{-0.25cm}
  \centering
  \includegraphics[width=.8\linewidth]{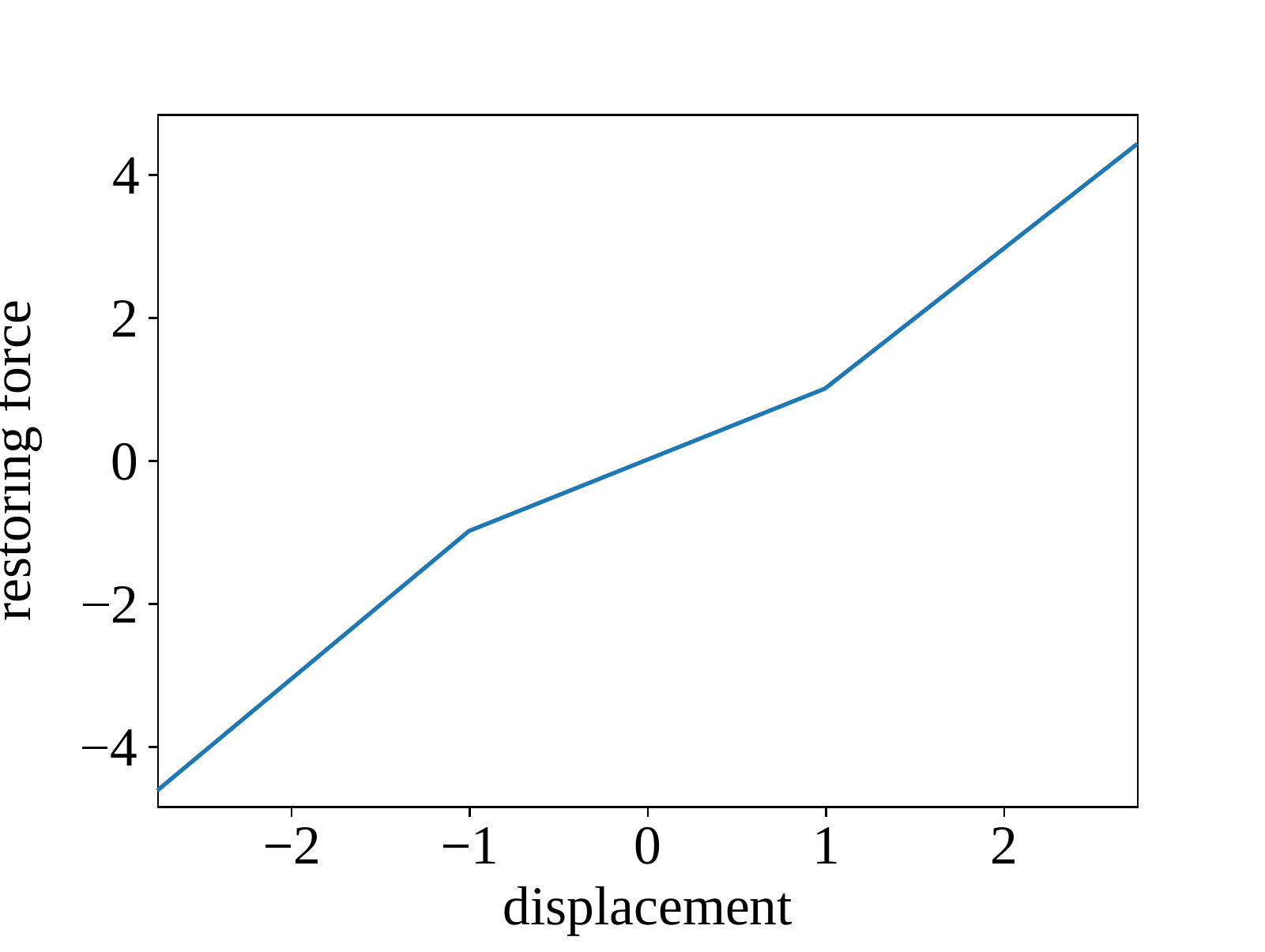}
  \label{fig:sub2}
\end{subfigure}
\caption{Panel (a) nonlinear forces due to the stoppers; Panel (b) total force-displacement relationship for a single mass, ignoring the coupling.}
\label{fig:NLforce}
\end{figure}

The model system under study is depicted in Fig. \ref{Fig:PhySys}. It consists of two masses $m$ coupled to the ground by identical springs $k_l$ and viscous dampers $c$. The masses are coupled to each other by a coupling spring $k_c$.
In the subsequent experimental setup we will employ coupled mechanical beams with bilateral contacts as the oscillators and describe there how the parameters can be determined. \\
To excite the system we assume that the ground is connected to a moving base with given periodic displacement $y(t)$. We also assume that there is no dissipation of energy when the oscillators touch the stoppers. Denoting the horizontal displacement of each oscillator with $x_1$ and $x_2$ , the dynamical system can be expressed as 
\begin{eqnarray}\label{eq:equationmotiontwodof}
m\ddot{x}_1 + c\dot{x}_1 + (k_l + k_c) x_1  - k_c x_2 + f_{nl}(x_1) = c \dot{y} + k_l y,\\
m\ddot{x}_2 + c\dot{x}_2 + (k_l + k_c) x_2  - k_c x_1 + f_{nl}(x_2) = c \dot{y} + k_l y,
\end{eqnarray}
where $f_{nl}$ represents the non-linear force, depicted in Fig.\ref{fig:NLforce}.
The non-linear force can be expressed as a piece-wise linear function, 
\begin{equation}
f_{nl}(x_i) = \begin{cases}
0 & \text{ if } |x_i| < s \\
    k_{nl}*(x_i-s) & \text{ if } x_i >s  \\
    k_{nl}*(x_i+s) & \text{ if } x_i <-s  \\
\end{cases}
\end{equation}
where $s$ might be called the gap dimension, characterising the amplitude that the oscillators need to reach before touching the stoppers.
\subsection{Numerical tools}
To compute periodic solutions of the numerical model, the harmonic balance method is applied  \cite{kerschen2009nonlinear, detroux2015harmonic}. The solution is determined in the form of a truncated Fourier series:
\begin{equation}
{\bf x}(t) = {\bf A}_0 + \sum_{k=1}^H {\bf A}_k \cos k \omega t + {\bf B}_k \sin k\omega t\;.
\end{equation} 
Substituting the expression into the equation of motion results in a nonlinear algebraic system of equations for the coefficients of the series. For a given frequency, the algebraic system can be solved using a Newton like root-finding algorithm. Continuation tools \cite{champneys1996numerical, doedel1998auto97, peeters2009nonlinear} allow to follow the results over varying parameters, like for example the frequency of the base excitation.  
\subsection{Linear and nonlinear modal analysis of the free conservative system}
\begin{figure}
	\begin{center}
		\vspace{-0.1cm}
		\includegraphics[]{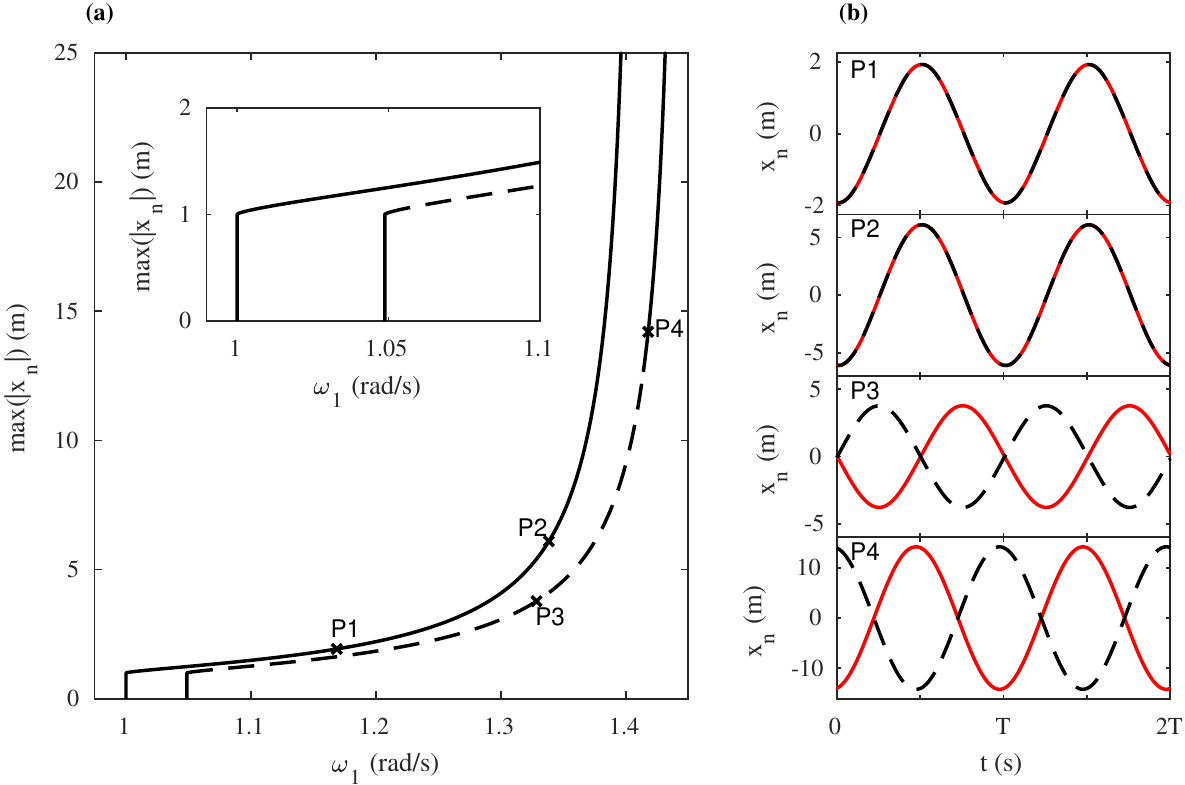}
		\caption{Backbone curves for the conservative nonlinear system, starting from the two linear normal modes. Panel (a) shows the bifurcation diagram, with full lines indicating stable solutions, while dashed lines represents unstable ones. Panel (b) shows the solutions identified in Panel (a) in time domain over two periods. The red lines denote $x_1$, the black lines denote $x_2$.}
		\label{fig:impactNNM1}
	\end{center}
\end{figure}

\begin{figure}
	\begin{center}
		\vspace{-0.1cm}
		\includegraphics[]{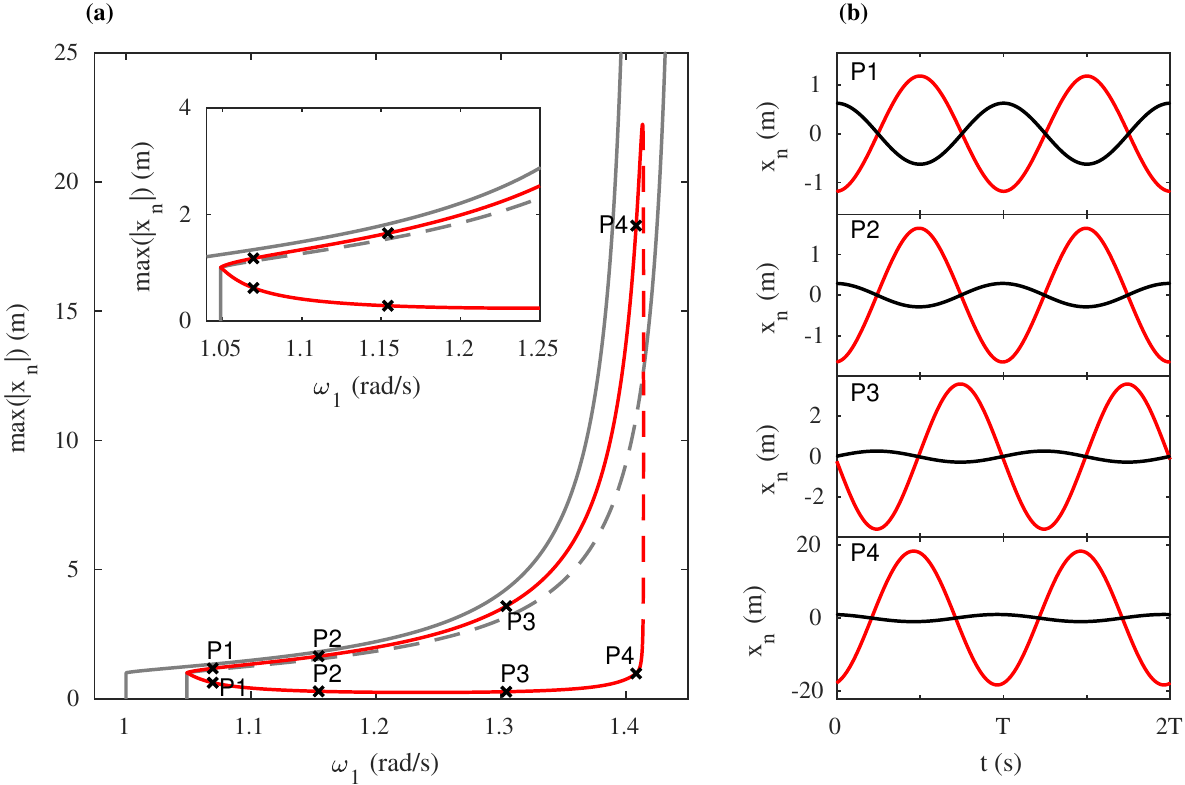}
		\caption{Backbone curves and bifurcating solution branch. Panel (a) depicts the solutions starting from the linear modes in grey and the bifurcating branch in red. The two red lines correspond to one new bifurcating solution branch and denote $x_1$ and $x_2$ respectively. Full lines denote stability, dashed ones instability. Panel (b) depicts the time dependency of solutions on the bifurcating branch of localised vibration.}
		\label{fig:impactNNM2}
	\end{center}
\end{figure}

We first study the linear and non-linear modes of the conservative form of the nonlinear two-degree-of-freedom system. The nonlinear modes are the periodic solutions of the undamped and unforced equations of motion,
\begin{eqnarray}\label{eq:equationmotiontwodof_free}
m\ddot{x}_1 + (k_l + k_c) x_1  - k_c x_2 + f_{nl}(x_1) = 0,\\
m\ddot{x}_2 + (k_l + k_c) x_2  - k_c x_1 + f_{nl}(x_2) = 0.
\end{eqnarray}
In the linear case, i.e. for low vibration amplitudes, there are two modes. One mode where the masses move in phase with equal amplitude, at frequency $\omega_1^2 = \frac{k_l}{m} $, and one mode where the masses move out of phase with equal amplitude, at a slightly larger frequency $\omega_2^2 = \frac{k_l + 2 k_c}{m}$ due to the coupling spring being activated. In the nonlinear case the analysis is carried out numerically. \\
In the following we assume $m=1$, $k_l=1$, $k_{nl}=1$, $s=1$, $k_c=0.05$, and $c=0.005$ where applicable. Typical results are presented in Fig.\ref{fig:impactNNM1}. The homogeneous in-phase and the out-of-phase modes continue to exist in the nonlinear regime, but their resonance frequencies depend on the amplitude of vibration. At amplitudes lower than the gap size, the modes are the modes of the linear system. At higher amplitudes, the oscillators start touching the stoppers and the frequencies increase due to the hardening-type nonlinearity. At very high amplitudes, the gap size becomes negligible compared to the vibration amplitude and the eigenfrequencies asymptotically approach those of the linear system with zero gap.\\
The stability analysis of the nonlinear normal modes, or periodic solutions, shows that the branch of the in-phase oscillations turns out to be linearly stable for all amplitudes. The out-of-phase oscillation, however, is linearly stable only in the linear regime, and again for larger vibration amplitudes, i.e. there is an amplitude range for which it is linearly unstable. It also turns out that in fact qualitatively new solutions bifurcate from the out-of-phase oscillations at the points of change of stability. Fig.\ref{fig:impactNNM2} shows the bifurcating solutions. The most remarkable property of the new solution class is that an extremely strong localisation of vibration amplitudes onto just one of the two oscillators is realised. \\
\subsection{Forced response}
\begin{figure}
	\begin{center}
		\vspace{-0.1cm}
		\includegraphics[]{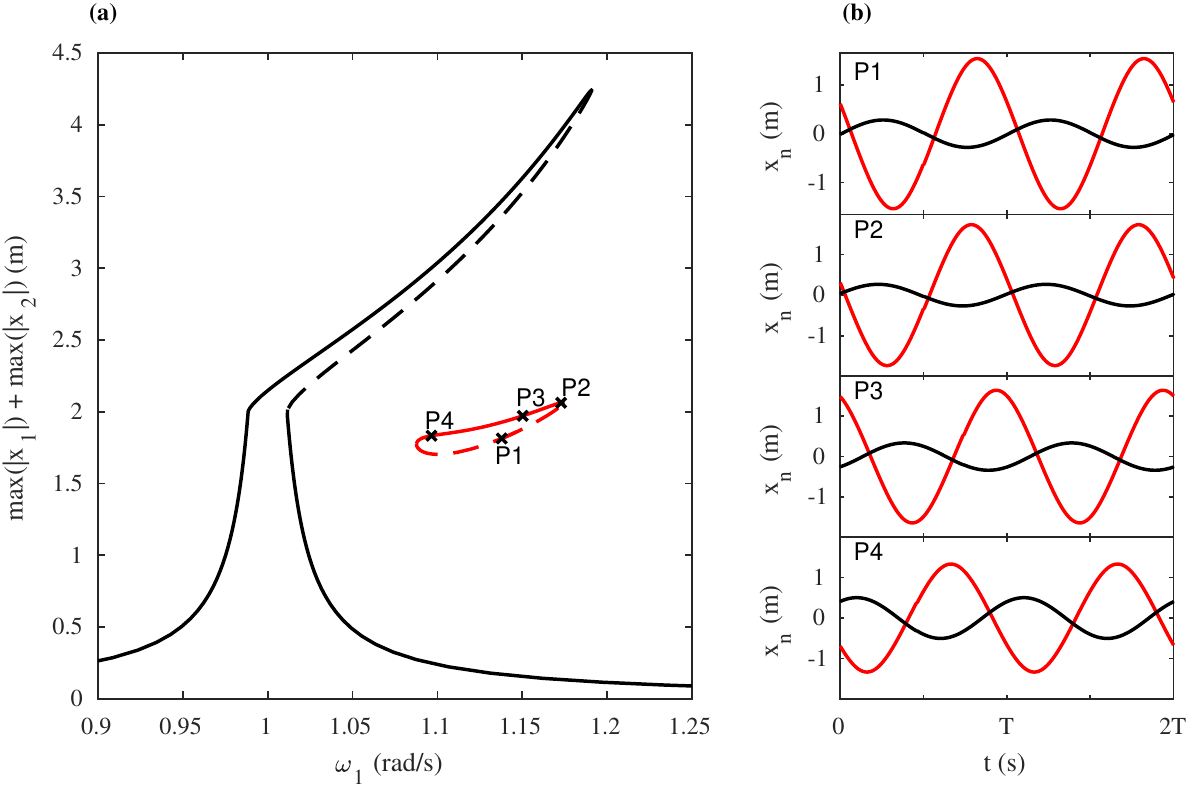}
		\caption{Forced response of the system in assuming a harmonic base excitation of amplitude $Y_0=0.025$. In Panel (a) the black lines are the homogeneous in-phase states, i.e. solutions with the masses vibrating in-phase and with the same amplitude. The red lines identify the isola of localised out-of-phase solutions. The solid lines are stable solutions, the dashed ones are unstable ones. Panel (b) depicts the time evolution of the two masses along two periods for the solutions identified in Panel (a).}
		\label{fig:impaceFRFF1}
	\end{center}
\end{figure}

\begin{figure}
	\begin{center}
		\vspace{-0.1cm}
		\includegraphics[]{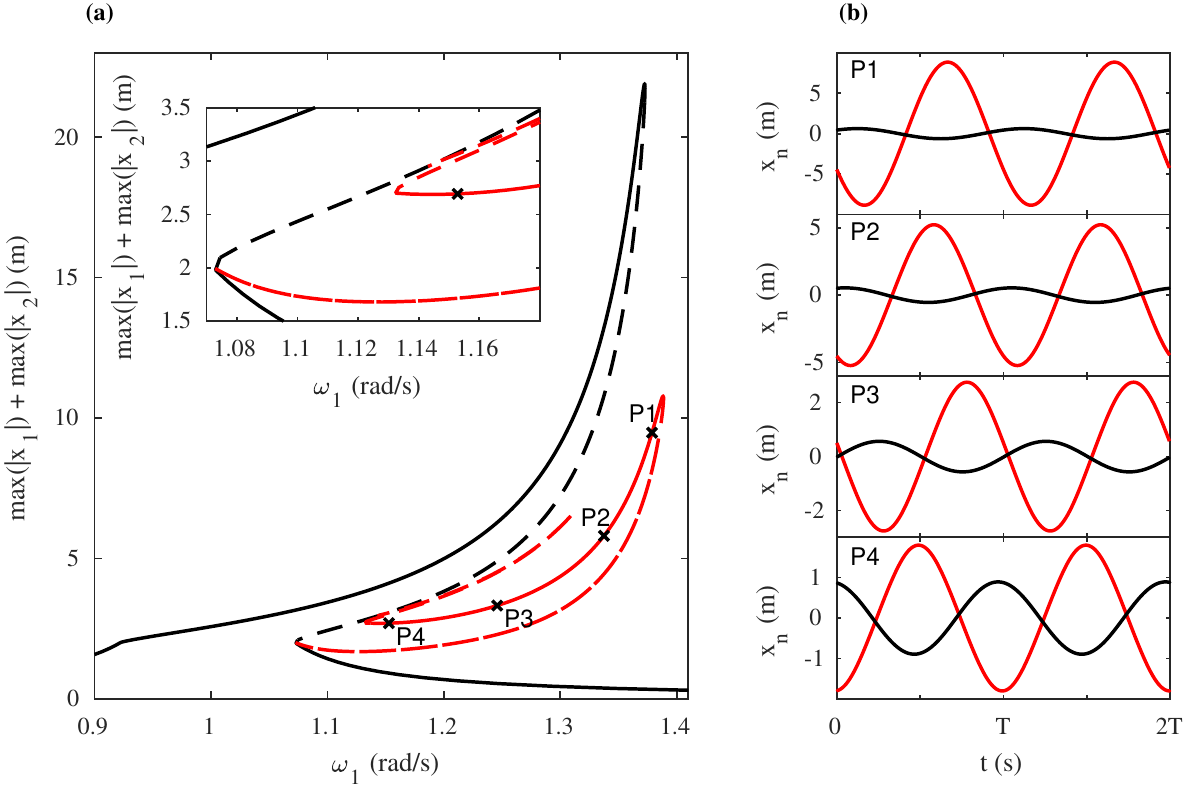}
		\caption{Forced response of the system assuming a harmonic base excitation of amplitude $Y_0 =0.15$. The two panels show the same quantities as in the previous figure.}
		\label{fig:impactFRFF3}
	\end{center}
\end{figure}

We now apply excitation in the form of a harmonically moving base, $y(t)=Y_0 \cos(\omega t)$, where $Y_0$ represents the amplitude of $y(t)$. First the system is studied assuming $Y_0=0.025$, which turns out to be large enough for the oscillators to reach the stoppers when the excitation frequency is near resonance. \\
Figure \ref{fig:impaceFRFF1} depicts the results. First one should note that due to the symmetry of the base excitation, from a linear systems perspective, out-of-phase solutions should not be excited, but merely in-phase solutions. Correspondingly, the response function of the in-phase mode shows the typical non-smooth stiffening behaviour for large response amplitudes, including its stability characteristics. However, out-of-phase solutions do show up in the form of an isola. These solutions can be obtained in the numerical approach by properly setting initial conditions for the root-finding process. The observed out-of-phase solutions are, however, not symmetric, but strongly asymmetric, or in other words localised on a single oscillator. As in the case of the free system, the out-of-phase solutions are related to localisation of the vibration amplitude onto one of the two oscillators. \\
One might note that the excitation of the localised solutions on the isola can be understood in terms of symmetry considerations. In the underlying linear regime, where the excitation is perfectly orthogonal to the out-of-phase mode, just the in-phase mode is triggered. This property also extends into the nonlinear range, the in-phase nonlinear normal mode is still symmetric, and thus excited, while the nonlinear out-of-phase normal mode remains fully anti-symmetric and is thus not excited. However, the localised modes bifurcating from the out-of-phase mode are not anti-symmetric any more. The projection of the symmetric external excitation onto these solution is thus non-vanishing and they can be excited. \\
The appearance of localisation through bifurcations and isola makes it plausible to expect that also the amplitude of the base excitation plays a decisive role. Fig. \ref{fig:impactFRFF3} depicts results obtained from a stronger base excitation of $Y_0=0.15$. Now the isola observed earlier has grown in size and the resulting branches of localised solutions merge back to the homogeneous in-phase solution.\\
\section{Experimental investigation}\label{Sec:Exps}
\subsection{Test setup}
\begin{figure}
		\centering
		\includegraphics[scale=0.2]{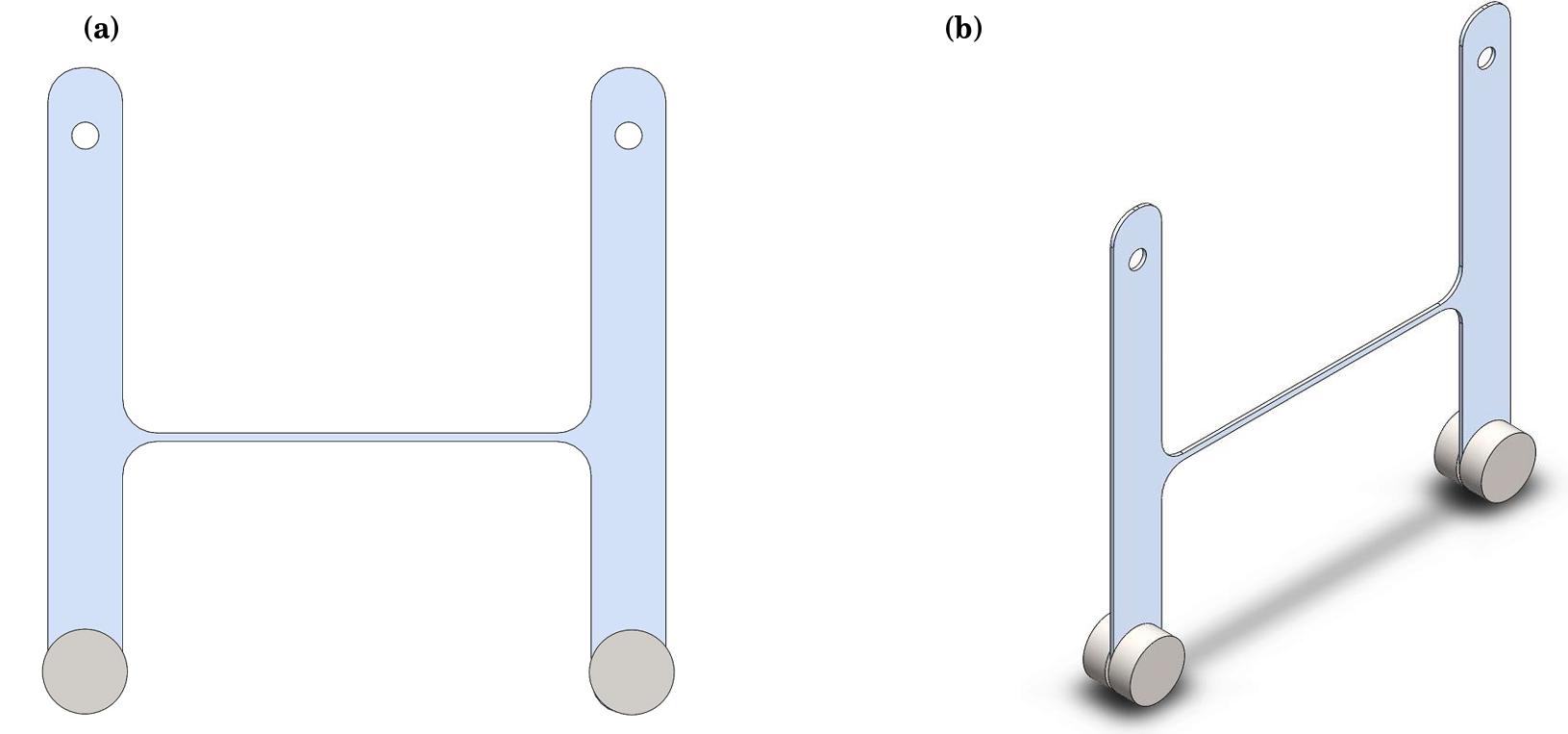}
		\caption{Structure designed to represent a system with two degrees of freedom. Panel (a) depicts a front view of the two beams with added masses at the tips. Panel (b) depicts an isometric illustration.}
		\label{fig:structure}
\end{figure}

\begin{figure}
	\begin{center}
		\vspace{-0.1cm}
		\includegraphics[width=\textwidth]{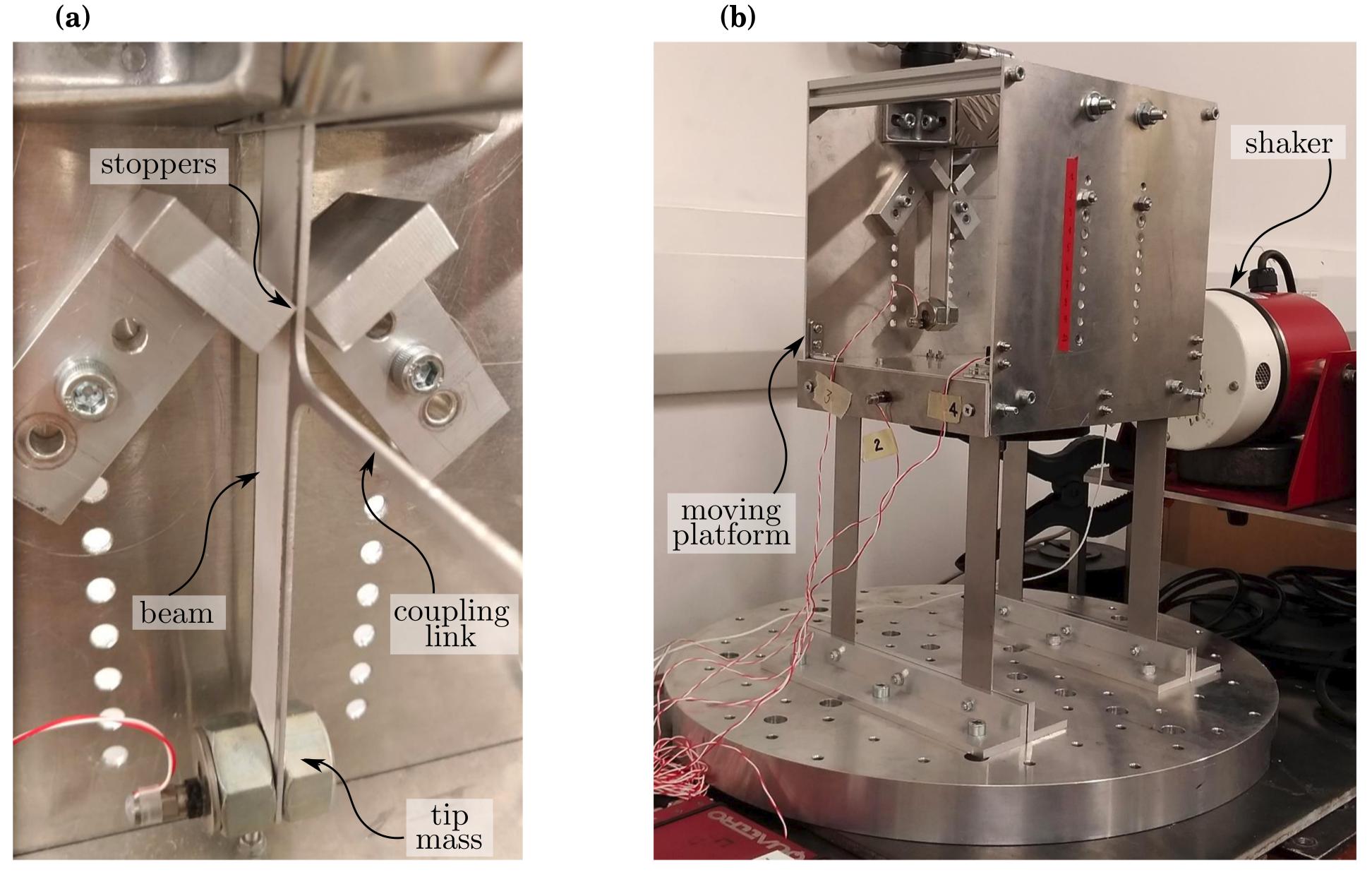}
		\caption{Test rig. Panel (a) depicts the stoppers near one of the two beams, while Panel (b) shows the platform connected to a shaker.}
		\label{fig:rig}
	\end{center}
\end{figure}

The experimental setup consists of two beams coupled to each other by a slender connection. The structure has been wire-cut from a single 1.5 mm thick aluminium sheet. A concentrated mass of approximately 70 g has been glued to the tip of each beam in order to force the beams to vibrate along their first bending mode, and also so that the mass of the beams becomes negligible and the beams behave as springs, see Figure \ref{fig:structure}.\\
The system is designed to show weak coupling between the two oscillators. In the experimental realisation the coupling is fully controlled by the slender structure connecting the two beams. If the connection is wider the structure becomes stiffer and the coupling stronger. A similar effect is obtained if the position of the slender connection is changed. If the connecting structure is positioned towards the tip of the beam, where displacements are relatively high, the coupling is also increased. In practice, the connection has been designed to obtain coupling values of a few percent of the beam stiffness.\\
The nonlinear effect is obtained by means of the beams contacting stoppers for large deflections. When reaching a certain vibration level, the beams start to touch the stoppers, see \mbox{Panel (a)} in Fig. \ref{fig:rig}, the effective beam length decreases due to the change in boundary conditions and the effective stiffness of the system increases.\\
The level of nonlinearity is controlled by the position of the stoppers. If the stoppers are moved towards the clamping position, the equivalent bending stiffness before
and after touching the stoppers are not so different. The opposite effect is obtained by moving the stoppers towards the tips of the beams. 
The base excitation, as assumed in the minimal model, is implemented by means of a moving platform. The two beams are clamped to a relatively rigid frame, and the final assembly is connected to the walls of the platform. A shaker is attached
to the platform, and the two beams are excited indirectly through the moving platform. \\
Panel (b) in Figure \ref{fig:rig} shows the test structure attached to the platform where an accelerometer is attached to the tip of the beam for measurements. The two stoppers near the blades are also illustrated. An impulse response test applying hammer excitation in the linear regime yields the in-phase and the out-of-phase natural frequencies for the test structure as \mbox{$11.18$ Hz} and \mbox{$11.62$ Hz.} These two values can be used to estimate an actual equivalent coupling ratio $k_c/k_l$ in the test rig of about $4 \%$.
\subsection{Test results and measurements}
\begin{figure}
	\begin{center}
		\vspace{-0.1cm}
		\includegraphics[]{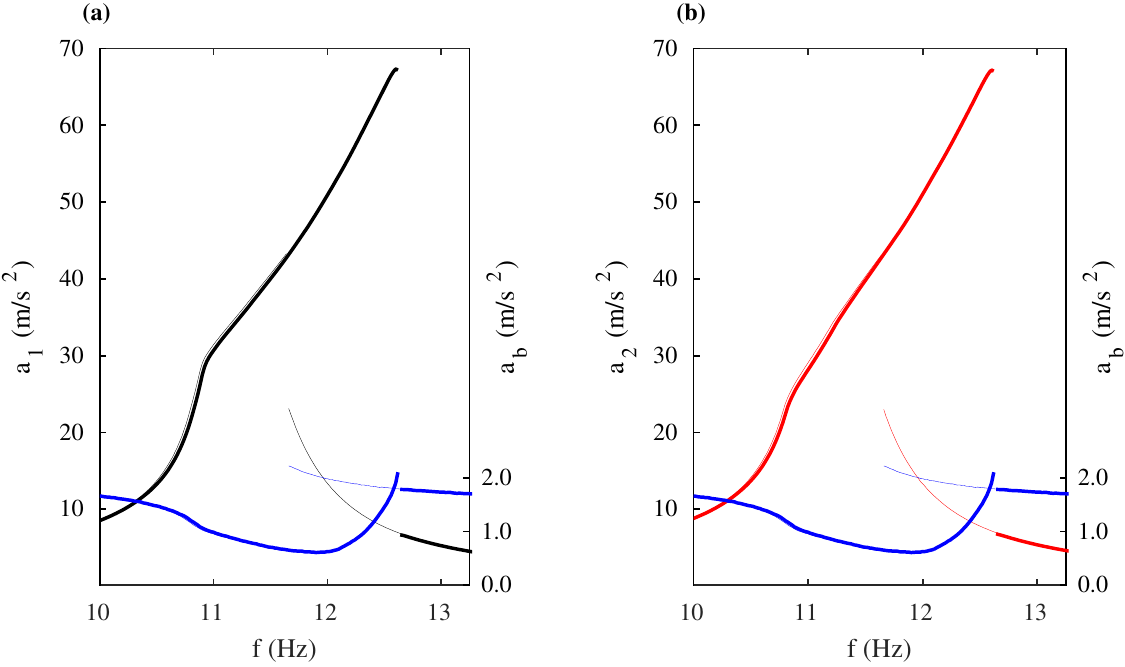}
		\caption{Response measured in the nonlinear regime. Panel (a) displays the acceleration of a single beam, where the thick black lines represent the measurements when the frequency is varied upward, while the thin lines represent the same results obtained when frequency is varied downward. The blue lines indicate the base acceleration. Panel (b) shows the same quantities as in Panel (a), but measured at the neighbouring beam.}
		\label{fig:FRFnlin_hom}
	\end{center}
\end{figure}

\begin{figure}
	\begin{center}
		\vspace{-0.1cm}
		\includegraphics[]{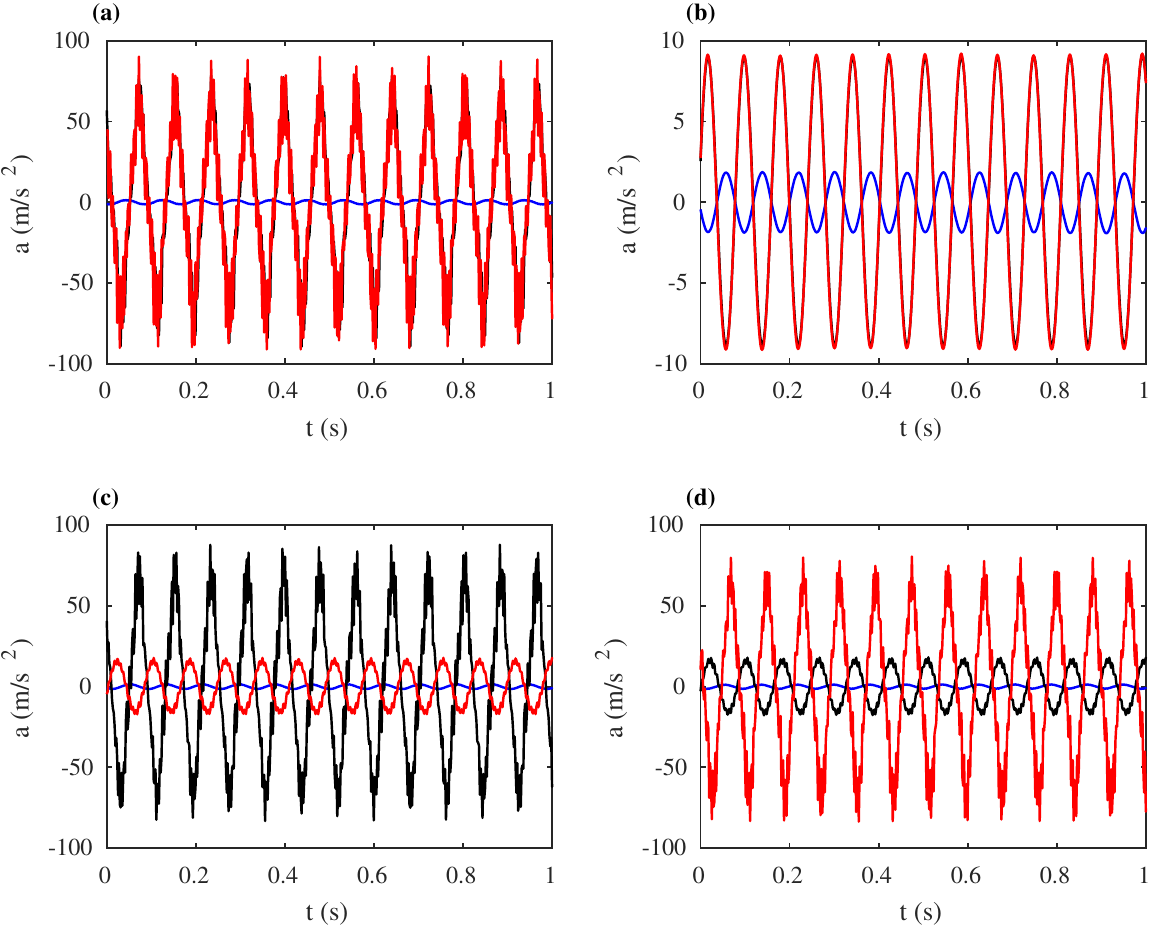}
		\caption{Accelerations measured in time-domain. Panel (a) depicts the homogeneous large amplitude state where both beams touch the stoppers. Panel (b) displays the small amplitude response, where both beams do not touch the stoppers. Panel (c) depicts the localised state when one of the beams touches the stoppers, the other one doesn't. Panel (d) displays the same behaviour, just for the localisation on the other beam. All panels show in black measurements for the first beam, while the red lines are for the second one. The results in blue depict the base acceleration.}
		\label{fig:NL_loctime}
	\end{center}
\end{figure}

\begin{figure}
	\begin{center}
		\vspace{-0.1cm}
		\includegraphics[]{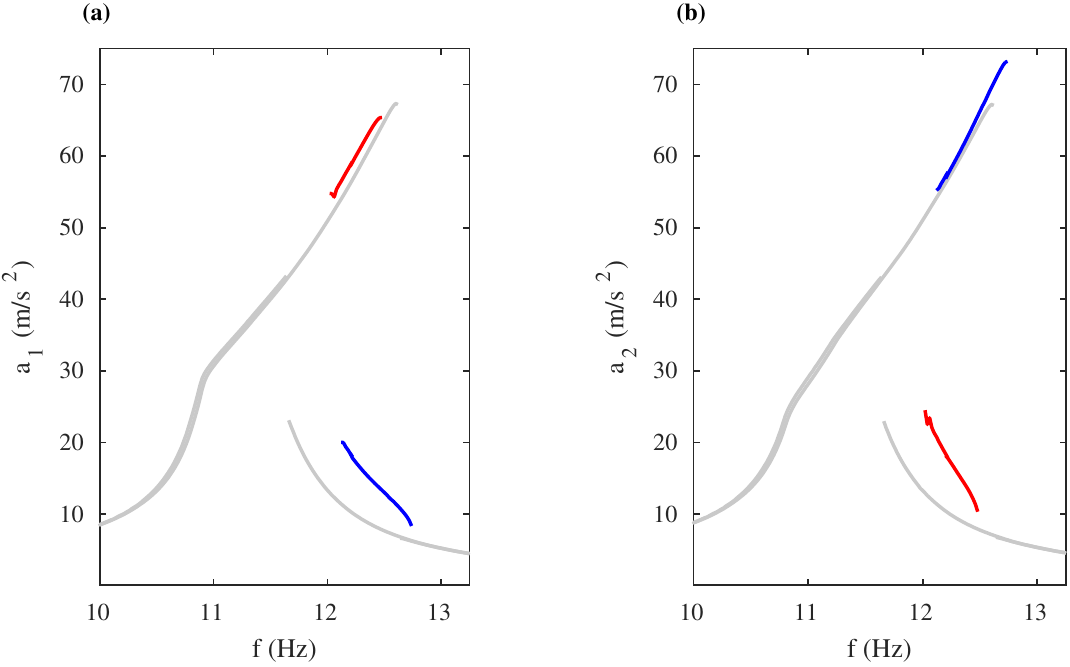}
		\caption{The two localised solutions, in red and blue lines respectively, identified experimentally. Panel (a) displays the response measured for the first oscillator, while Panel (b) depicts the response measured for the second one. The grey lines illustrate the homogeneous in-phase solutions.}
		\label{fig:FRFnlin_loc}
	\end{center}
\end{figure}

The structure is first tested in the linear regime where the imposed force produced by the shaker is not large enough to drive the two beams into contact with the stoppers. The motivation of this test is to investigate the level of remaining inhomogeneities.
If the system is perfectly symmetric, the response measurements should depict a single resonance frequency only, since the out-of-phase mode should not be excited. The acceleration is measured at the tip of each beam and also at the platform. \\
\begin{figure}
	\begin{center}
		\vspace{-0.1cm}
		\includegraphics[]{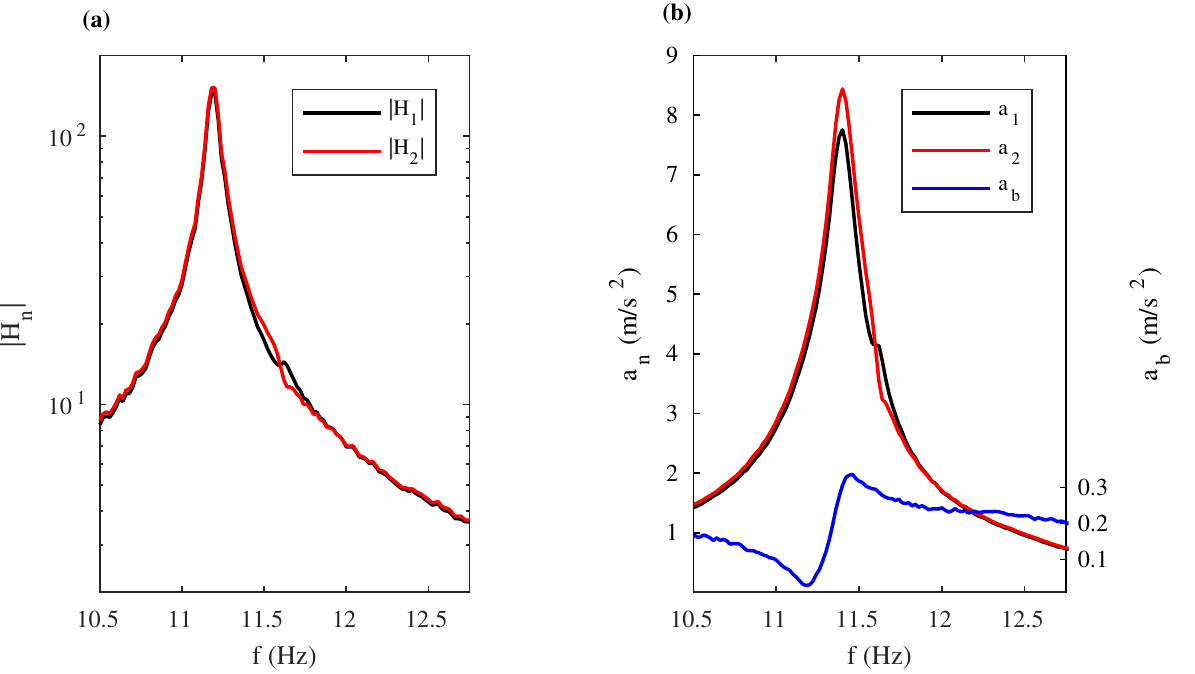}
		\caption{Response measured in the linear regime. Panel (a) displays the resulting amplitude response ratio $|H_n | = |a_n |/|a_b|$, while Panel (b) displays the measured accelerations directly.}
		\label{fig:FRFlinear}
	\end{center}
\end{figure}
Panel (a) of Fig. \ref{fig:FRFlinear} depicts the response of the two beams with $H_n = {a_n}/{a_b}$, where $a_n$ is the measured acceleration for the respective beam and $a_b$ is the corresponding acceleration measured at the platform. The results are plotted in logarithmic scale. The transmissibility displays a single peak centred at $f = 11.18$ Hz, with respective modal damping of $\xi = 0.55\%$. A very small contribution from the out-of-phase mode can, due to remaining inhomogeneities, be identified around $f =11.62$ Hz. For completeness, Fig \ref{fig:FRFlinear} also displays the directly measured results in a linear scale. \\
For stronger base excitation, the beams start touching the stoppers and the system becomes nonlinear. As to be expected from the modelling results, hysteresis effects are to be observed and the excitation is varied upwards and downwards in frequency. Figure \ref{fig:FRFnlin_hom} displays the corresponding measured results. \\
The graph in Panel (a) depicts the acceleration measured at one of the beams, while Panel (b) depicts the same quantities for the neighbouring oscillator. In both panels the blue lines indicate the measurements for the base. The measurements for the first beam, in Panel (a), show in thick black lines the results when the excitation frequency is increased. The system follows the upper branch of homogeneous in-phase
solutions and jumps back to the low-amplitude configuration at $f \sim12.5$ Hz. The results obtained when the frequency is decreased are displayed in thin black lines. In this case, the system jumps to large-amplitude states only near the point of initial contact when the oscillator starts touching the stoppers at $f  \sim11.75$ Hz. The solutions linking the two turning points are not measured due to their linear instability. The results in Panel (b) follow the same measurement strategy, but are measured on the other beam. The results are in excellent agreement, confirming the symmetry of the setup. The results are also in excellent agreement with the response characteristics of the nonlinear model treated earlier. \\
Modelling and simulation also predict that a branch of localised solutions may exist. The branches leading to localised vibrations arise either through bifurcations from the homogeneous branch or in the form of isolas, depending on the excitation level. In order to test the existence of such kind of a nonlinear vibration localisation in the experiment, appropriate initial conditions have been employed, consisting in triggering high amplitude vibration in one of the oscillator. 
The platform was excited at $f =12.3$ Hz, the system settled into the low-amplitude in-phase state, and only then one of the beams was pulled towards the stoppers by external means. The main motivation was to introduce a perturbation which forces the nonlinear dynamical system to jump from the observed homogeneous in-phase branch to the desired localised one, whose existance was expected in this frequency range. Figure \ref{fig:NL_loctime} displays the four stable stationary response configurations observed in the time-domain for a base excitation at $f =12.3$ Hz.\\
Panels (a) and (b) of Fig. \ref{fig:NL_loctime} refer to the states already measured, they correspond to the large and small amplitude in-phase response, respectively. In Panel (a) both beams vibrate in-phase and touch the stoppers, with some noticeable high-frequency components induced by the contacts. Panel (b) shows the small amplitude linear case where both beams do not touch the stoppers and the measured accelerations are nearly sinusoidal. \\
The first regime of localised vibrations observed experimentally is indicated in Panel (c). Within this configuration only the first beam touches the stopper and keeps vibrating in a nonlinear large amplitude manner, while the neighbouring oscillator remains in low amplitude, not touching the stoppers. The localised solution measured in the time-domain is in excellent agreement with the results predicted numerically. \\
Due to the symmetry inherent in the system, also the analogous vibration localisation on the other oscillator is expected. To test this hypothesis of the existence of the symmetric state, the neighbouring beam is now pulled towards the stopper. Panel (d) of Fig. \ref{fig:NL_loctime} shows the
corresponding measured response. Obviously the nonlinear vibration localisation can thus also arise on the other oscillator. It depends merely on the choice of initial condition where the localisation will happen. In the supplementary online-material of this paper a video illustrates all states discussed. \\
One should note that in the experimental setup, for the localised state the acceleration values measured for the beam vibrating in large amplitudes exceed four times the same quantities for the neighbouring oscillator in small amplitude. Moreover, the observed state seems to be very robust, and the system remains in the localised branch even if the beams are slightly perturbed externally. \\
Once one localised state is reached, the whole stable branch of localised states can be traced experimentally. The approach is similar to the experimental strategy implemented to measure the branch of homogeneous, i.e. in-phase solutions. First, the system is perturbed externally
until the desired localised configuration is achieved. Then the excitation frequency is varied
in frequency until the physical system jumps back from the localised branch to the underlying
homogeneous one. This transition delineates the range of existence of the stable localised solutions at hand. The same approach can be carried out varying the excitation frequency downwards. The measured
localised responses trace the underlying branch of stable non-homogeneous solutions. Figure \ref{fig:FRFnlin_loc} depicts the experimental results. \\
The results depicted in blue and red lines denote the two localised solutions. One may note that when e.g. the first beam vibrates with touching the stoppers, as depicted by the large
amplitude of the red line in Panel (a), the corresponding neighbour oscillator is in low amplitude,
as illustrated by the red line in Panel (b). And vice versa for localisation on the second beam. Ideally, the measured response in Panels (a) and (b) should be identical, which is the case to a surprisingly good extent, considering the usual technical difficulties in measuring nonlinear mechanical vibrations. 
\section{Summary, conclusions and outlook}\label{Sec:Conc}
This work focused on the investigation of a symmetric system with piecewise
nonlinearity. A minimal model with two degrees of freedom was set up to study the properties of localised vibration due to nonlinearity. An experimental setup confirmed and validated the model-based findings. \\
First a conservative analysis was carried out in order to compute the nonlinear normal modes
of the model system. The underlying linear regime shows the usual normal modes,
where the two oscillators vibrate homogeneously in phase or out of phase. The nonlinear analysis showed that novel states emerge bifurcating from the out-of-phase mode, characterised by vibration localisation on single oscillators. In frequency ranges near the linear resonance, the out-of-phase homogeneous states may result unstable and the localised response is stable instead. \\
The nonlinear localised states also exist in the driven case. The analysis was performed
assuming a harmonic base excitation. The underlying linear analysis suggests that only
the in-phase mode should be excited since the external forcing and the out-of-phase
mode are perfectly orthogonal. However, due to the existence of the bifurcated, asymmetric localised modes, the nonlinear localised modes can be triggered through the base excitation. Consequently,
localised vibrations, arising from the nonlinear system dynamics, may easily result when the nonlinear regime is reached. Depending on the level of external forces they arise in the form of isolas or
through bifurcations. Moreover, large parts of the branches of localised states turn out as linearly stable and are so to be expected to be observed directly. \\
In order to test and validate the findings from the model, a test setup was designed and built. The experimental system consisted of two weakly coupled
beams with the effect of frequency changing contacts for large vibration amplitudes. The configuration was deliberately set up to reproduce the bilinear stiffness behaviour of the model. Each beam vibrates as a simple oscillator with effectively softer or stiffer springs, depending on the vibration amplitude. Harmonic base excitation of the platform was applied. \\
The system was tested and analysed in the linear and nonlinear regimes. The response in the linear case shows the expected single-mode excitation, i.e. response in the form of a symmetric in-phase mode. For large amplitudes, when the beams start touching the stoppers and thus the system becomes
nonlinear, localised response can be observed. Within the given configuration the system localises vibrations in either the first or the second beam, depending on the initial
conditions. Depending on the frequency and the level of excitation, four stable
configurations may arise and have been observed experimentally. In the first one, both beams touch
the stoppers, resulting in a nonlinear homogeneous state. In the second configuration,
none of the beams touches the stoppers, leading to a homogeneous and purely linear response.
In the two other configurations the localised states are induced by nonlinearity, and vibration localises in either the first of the second beam. \\
The present work has attempted to contribute to the understanding of vibration localisation in symmetric systems caused by nonlinearity. Both modelling and testing show conclusive evidence that the studied symmetric, weakly coupled two degree-of-freedom oscillator may tend to respond in the form of localised vibrations when the driving is near the resonance of the individual oscillators, and when the forcing amplitude is strong enough. \\
The findings have close analogies with results in nonlinear localisation from various fields of physics \cite{flach1998discrete, flach2008discrete}. The present work might thus be considered an attempt to better understand the mechanics of nonlinear vibration localisation in nonlinear structures of engineering. Future work will need to clarify further aspects. Amongst others, the role of the vibration amplitudes necessary to reach the effects in actual applications, the role of nonlinear vibration localisation in larger systems with more than two degrees of freedom, the interplay between mistuning related and nonlinear vibration localisation. 
\begin{acknowledgements}
	The first author is funded by the Brazilian National Council for the Development of Science and Technology (CNPq) under the grant 01339/2015-3.
\end{acknowledgements}

%
\section*{Conflict of interest}

The authors declare that they have no conflict of interest.

\bibliographystyle{spmpsci}      
\bibliography{mybibfile3}

\end{document}